\def\be{\begin{equation}}\def\ee{\end{equation}}
\def\bea{\begin{eqnarray}}\def\eea{\end{eqnarray}}
\def\bd{\begin{displaymath}}\def\ed{\end{displaymath}}

\def\hx{\hat x}\def\hp{\hat p}\def\xp{\,x{\cdot}p\,} \def\px{\,p{\cdot}x\,}

\def\cH{{\cal H}}\def\cA{{\cal A}}\def\cC{{\cal C}}
\def\lra{\leftrightarrow}
\def\ha{{1\over2}}\def\qu{{1\over4}}

\def\PL#1{Phys.\ Lett.\ {\bf#1}}\def\CMP#1{Commun.\ Math.\ Phys.\ {\bf#1}}
\def\PRL#1{Phys.\ Rev.\ Lett.\ {\bf#1}}
\def\PR#1{Phys.\ Rev.\ {\bf#1}}\def\CQG#1{Class.\ Quantum Grav.\ {\bf#1}}
\def\NP#1{Nucl.\ Phys.\ {\bf#1}}

\def\JoP#1{J.\ Phys.\ {\bf#1}} \def\IJMP#1{Int.\ J. Mod.\ Phys.\ {\bf #1}}

\def\JHEP#1{JHEP\ {\bf#1}}\def\JCAP#1{JCAP\ {\bf#1}}\def\EPJ#1{Eur.\ Phys.\ J.\ {\bf#1}}
\def\RMP#1{Rev.\ Mod.\ Phys.\ {\bf#1}}\def\AdP#1{Annalen Phys.\ {\bf#1}}
\def\AHEP#1{Adv.\ High En.\ Phys.\ {\bf#1}}
\def\hep#1{{\tt hep-th/#1}}

\def\JETP#1{JETP\ {\bf#1}}

\documentclass[a4paper,superscriptaddress,nofootinbib,10pt]{article}
\usepackage{geometry}
\begin{document}

{
\vskip40pt
\centerline{\bf  Noncommutative Yang model and its generalizations}
\vskip40pt
\centerline{{\bf S. Meljanac}\footnote{e-mail: meljanac@irb.hr}}
\vskip5pt
\centerline {Rudjer Bo\v skovi\'c Institute, Theoretical Physics Division}
\centerline{Bljeni\v cka c. 54, 10002 Zagreb, Croatia}
\vskip10pt
\centerline{and}
\vskip5pt
\centerline{{\bf S. Mignemi}\footnote{e-mail: smignemi@unica.it}}
\vskip5pt
\centerline {Dipartimento di Matematica, Universit\`a di Cagliari}
\centerline{via Ospedale 72, 09124 Cagliari, Italy}
\smallskip
\centerline{and INFN, Sezione di Cagliari}
\centerline{Cittadella Universitaria, 09042 Monserrato, Italy}

\vskip80pt
\centerline{\bf Abstract}
\medskip
{\noindent
Long time ago, C.N. Yang proposed a model of noncommutative spacetime that generalized the Snyder model to a curved background.
In this paper we review his proposal and the generalizations that have been suggested during the years. In particular, we discuss
the most general algebras that contain as subalgebras both de Sitter and Snyder algebras, preserving Lorentz invariance, and
are generated by a two-parameter deformation of the canonical Heisenberg algebra.
We also define their realizations on quantum phase space, giving explicit examples, both exact and in terms of a perturbative
expansion in the deformation parameters.
}
\vskip60pt
\vfil\eject}

\section{Introduction}
At the Planck scale, gravitational and quantum effects have comparable strength and can affect the structure of spacetime.
In this regime, noncommutative geometries may play a relevant role \cite{1}.
The first example of noncommutative spacetime was proposed by Snyder\cite{2} in 1947, but at the time did not receive much attention.
In spite of the assumed granular structure of spacetime, this model is characterized by the preservation of the Lorentz invariance.

A first generalization of Snyder's idea was advanced by C.N.~Yang, who combined noncommutativity with spacetime curvature \cite{3}, in terms of
a fifteen-dimensional SO(1,5) algebra of symmetries of phase space. The generators of this algebra were identified with the generators of the de Sitter
algebra and with the coordinates of the de Sitter spacetime. The remaining generator $h$ rotates positions into momenta, but its physical interpretation was
not specified. More recently, this model was slightly generalized by Khruschev and Leznov (KL) \cite{4}.

Later, in ref.~\cite{5} was proposed a model inspired by that of Yang, which realizes the same symmetries in a nonlinear way, reducing to fourteen the number
of independent generators of the algebra, thus eliminating the unphysical generator $h$. This model was dubbed triply special relativity (TSR) because it contains
three fundamental constants, identified with the speed of light, the Planck mass and the cosmological constant, generalizing in this way the idea advanced
in doubly special relativity theories \cite{6} of deforming the Poincar\'e symmetry by the introduction of a new fundamental constant proportional to the Planck mass.
In this model, however, the Lorentz symmetry is preserved and only translation symmetries are deformed.

It was then shown in \cite{7} that the TSR model can be realized exactly in terms of coordinates and momenta only. This particular realization in phase space was called
Snyder-de Sitter (SdS) spacetime. In \cite{7,8} it was also shown that the SdS algebra can be obtained from the Snyder algebra by a nonunitary transformation.

The previous models of quantum phase space have the common property of realizing explicitly the duality introduced by Born \cite{9} for the exchange of position
and momentum operators.
Recently, the possibility of studying general quantum phase spaces displaying the structure described above has been advanced in \cite{10} and studied in more detail in \cite{11}.
These papers employed a widespread approach to noncommutative geometry based on the formalism Hopf algebras \cite{12}, that aims to describe the symmetries of the quantum spacetime.
A powerful tool in this formalism are the realizations of Hopf algebras in terms of the Heisenberg algebra, that were introduced in \cite{13,14,15}.

The Snyder model has been extensively discussed in the literature, see e.g.~\cite{13,14,16,17}, in several theoretical and phenomenological aspects, like algebraic
representations, relativistic field equations, quantum field theory, relations with Hopf algebras, but also its nonelativistic limit has been widely investigated,
including generalized uncertainty principles, simple models like the harmonic oscillator, etc. To our knowledge, instead, besides ref.~\cite{4}, the Yang model has been
considered only in \cite{18}, where its supersymmetric extensions were analyzed.
More authors discuss aspects of TSR, see e.g.~\cite{7,8,11,19}.  Most of them treat its classical limit, either in a nonrelativistic or relativistic setting, but
also the quantum field theory of a self-interacting scalar field in SdS spacetime has been investigated in \cite{20}.

In the present paper, we discuss general perturbative realizations of the unified model proposed in \cite{10}, in terms of the canonical Heisenberg algebra,
extending the results of ref.~\cite{11}. In sect.~2 we review the generalized Snyder spaces and their realizations.
In sect.~3 we discuss the de Sitter algebras  obtained by duality from the Snyder ones. In sect.~4 a formalism unifying the two algebras
is introduced, that describes a curved noncommutative spacetime and includes Yang, KL and TSR models as special cases. Sect.~5 analyzes the perturbative
realizations of these models, while in sect.~6 we exploit the relation of TSR with the Snyder model to write down some exact realizations.

\section{Generalized Snyder space}
In this section we review some generalizations of Snyder's original proposal for a deformation of the momentum space, that leads
to noncommutativity of spacetime.
Most of the results of this section were obtained in \cite{13,14,17}.

We define a generalized Snyder space introducing a Lorentz-invariant deformation parameter $\beta\sim L_P$, and postulate the commutation relations
\be\label{e1}
[\hx_\mu,\hx_\nu]=i\epsilon\beta^2M_{\mu\nu}\;\psi^p(\epsilon\beta^2p^2),\qquad[p_\mu,p_\nu]=0,\qquad[\hx_\mu,p_\nu]=i\phi_{\mu\nu}^p(\epsilon\beta^2p^2),
\ee
where $\epsilon=\pm1$ and $M_{\mu\nu}=M_{\mu\nu}^{\,\dagger}$ are the generators of the Lorentz algebra, that we assume to satisfy the standard relations,
\bea\label{e2}
&&[M_{\mu\nu},M_{\rho\sigma}]=i\big(\eta_{\mu\rho}M_{\nu\sigma}-\eta_{\mu\sigma}M_{\nu\rho}-\eta_{\nu\rho}M_{\mu\sigma}+\eta_{\nu\sigma}M_{\mu\rho}\big),\cr
&&[M_{\mu\nu},\hx_\lambda]=i(\eta_{\mu\lambda}\hx_\nu-\eta_{\nu\lambda}\hx_\mu),\qquad[M_{\mu\nu},p_\lambda]=i(\eta_{\mu\lambda}p_\nu-\eta_{\nu\lambda}p_\mu)
\eea
with $\eta_{\mu\nu}$ the flat metric, $\eta_{\mu\nu}={\rm diag}(-1,1,1,1)$, and the functions $\psi^p(\epsilon\beta^2p^2)$ and $\phi_{\mu\nu}^p(\epsilon\beta^2p^2)$ are
constrained so that the Jacobi identities hold.
The commutation relations (\ref{e1})-(\ref{e2}) generalize those originally investigated in \cite{2}, which are recovered for $\psi^p={\rm const}$ and
$\phi^p_{\mu\nu}=\eta_{\mu\nu}+\beta^2p_\mu p_\nu$.

In the following, we shall look for realizations of the algebra (\ref{e1})-(\ref{e2}) in terms of the Heisenberg algebra $\cH$:
in its undeformed version the Heisenberg algebra  is generated by commutative coordinates $x_\mu$ and momenta $p_\mu$ that
satisfy
\be
[x_\mu,x_\nu]=[p_\mu,p_\nu]=0,\qquad[x_\mu,p_\nu]=i\eta_{\mu\nu},
\ee
with $x_\mu^\dagger=x_\mu$ and $p_\mu^\dagger=p_\mu$.

The action of $x_\mu$ and $p_\mu$ on functions $f(x)$ belonging to the enveloping algebra $\cA$ generated by the $x_\mu$ is
defined as
\be
x_\mu\triangleright f(x)=x_\mu f(x),\qquad p_\mu\triangleright  f(x)=-i{\partial f(x)\over\partial x^\mu}.
\ee

Noncommutative coordinates $\hx_\mu$ and Lorentz generators $M_{\mu\nu}$ can be expressed in terms of commutative coordinates $x_\mu$ and
momenta $p_\mu$ as
\bea\label{e5}
&&\hx_\mu=x_\mu\varphi^p_1(\epsilon\beta^2p^2)+\epsilon\beta^2x{\cdot}p\,p_\mu\varphi^p_2(\epsilon\beta^2p^2)+\epsilon\beta^2p_\mu\chi^p(\epsilon\beta^2p^2),\cr
&&M_{\mu\nu}=x_\mu p_\nu-x_\nu p_\mu,\qquad M_{\mu\nu}=M^\dagger_{\mu\nu}.
\eea
In terms of the realizations (\ref{e5}), the functions $\phi_{\mu\nu}^p$ in (\ref{e1}) read
\be
\phi^p_{\mu\nu}=\eta_{\mu\nu}\varphi^p_1+\epsilon\beta^2p_\mu p_\nu\varphi^p_2.
\ee
The Jacobi identities are satisfied if
\be\label{e7}
\psi^p=-2\varphi^p_1(\varphi^p_1)'+\varphi^p_1\varphi^p_2-2\epsilon\beta^2p^2(\varphi^p_1)'\varphi^p_2,
\ee
where a prime denotes a derivative with respect to $\epsilon\beta^2p^2$.
The function $\psi^p$ is independent of $\chi^p$.

From (\ref{e7}) it follows that the $\hx_\mu$ are commutative for
\bd
\varphi^p_2={2\varphi^p_1(\varphi_1^p)'\over\varphi^p_1-2\epsilon\beta^2p^2(\varphi^p_1)'},
\ed
and correspond to Snyder space if
\bd
\varphi^p_2={1+2\varphi^p_1(\varphi_1^p)'\over\varphi_1^p-2\epsilon\beta^2p^2(\varphi_1^p)'}.
\ed

Some interesting special cases arise for particular values of the functions $\varphi_i^p$.
The Snyder realization \cite{2,14} is recovered for $\varphi^p_1=\varphi^p_2=1$ and the Maggiore realization \cite{21,14}
for $\varphi^p_1=\sqrt{1-\epsilon\beta^2p^2}$ and $\varphi^p_2=0$.
Another interesting case of realization of Snyder space \cite{17} is for $\varphi^p_1=s=$ const, $\chi^p=0$,
\be
\hx_\mu=x_\mu+{\epsilon\beta^2s\over4}K_\mu,
\ee
where $K_\mu=x_\mu p^2-2x{\cdot}p\,p_\mu$ are the generators of conformal transformations in momentum space, with
$[K_\mu,K_\nu]=0$.

The algebra (\ref{e1})-(\ref{e2}) unifies commutative space, $\psi^p=0$, and Snyder space, $\psi^p=1$. Since the Lorentz transformations are not deformed,
its Casimir operator is $\cC^p=p^2$.
The algebra (\ref{e1})-(\ref{e2}) is invariant under Hemitian conjugation if $p_\mu^\dagger=p_\mu$, $\hx_\mu^\dagger=\hx_\mu$,
$M_{\mu\nu}^\dagger=M_{\mu\nu}$.

Hermitian realizations of $\hx_\mu(\beta)$ are obtained from (\ref{e5}) as
\be\label{e9}
\hx_\mu(\beta)=\ha\Big[x_\mu\varphi^p_1+\varphi^p_1x_\mu+\epsilon\,\beta^2(x{\cdot}p\,p_\mu\varphi^p_2+\varphi^p_2p_\mu p{\cdot}x)\Big]+\epsilon\,\beta^2p_\mu\chi^p,
\ee
where $\varphi_1^p$, $\varphi_2^p$, and $\chi^p$ are real functions.

\section{Noncommutative momenta of generalized Snyder type}
By duality, one can obtain different Snyder-like algebras with noncommutative momenta, that coincide with generalized
de Sitter algebras and can be associated to the symmetries of spacetimes of constant curvature, namely de Sitter or anti-de Sitter.
The relevance of de Sitter spacetime in general relativity and quantum gravity is well-known \cite{22}, especially in cosmological context.
The algebra of symmetries of de Sitter spacetime is also well-known, see e.g.~\cite{23}.
Also deformations of special relativity based on the de Sitter algebra have been investigated \cite{24}.
The models presented here correspond to unusual parametrizations of the de Sitter manifold, which are isotropic in the 4-dimensional spacetime \cite{25}.

To obtain these algebras, we introduce a new Lorentz-invariant deformation parameter $\alpha$ of dimension inverse length, so that $\alpha^2$
may be identified with the cosmological constant, and assume
\be\label{e10}
[\hp_\mu,\hp_\nu]=i\epsilon'\alpha^2M_{\mu\nu}\psi^x(\epsilon'\alpha^2x^2),\qquad [x_\mu,x_\nu]=0,\qquad [x_\mu,\hp_\nu]=i\phi_{\mu\nu}^x(\epsilon'\alpha^2x^2),
\ee
with $\epsilon'=\pm1$. Moreover, the Lorentz generators satisfy
\be\label{e11} 
[M_{\mu\nu},\hp_\lambda]=i(\eta_{\mu\lambda}\hp_\nu-\eta_{\nu\lambda}\hp_\mu),
\ee
and the other standard relations in (\ref{e2}).
The functions $\psi^x(\epsilon'\alpha^2x^2)$ and $\phi_{\mu\nu}^x(\epsilon'\alpha^2x^2)$ are chosen so that the Jacobi identities are satisfied.

The noncommutative momenta $\hp_\mu$ and the Lorentz generators $M_{\mu\nu}$ can then be expressed in terms of commutative coordinates and momenta as
\bea\label{e12}
&&\hp_\mu=p_\mu\varphi^x_1(\epsilon'\alpha^2x^2)+\epsilon'\alpha^2p{\cdot}x\,x_\mu\varphi^x_2(\epsilon'\alpha^2x^2)+\epsilon'\alpha^2x_\mu\chi^x(\epsilon'\alpha^2x^2),\cr
&&M_{\mu\nu}=x_\mu p_\nu-x_\nu p_\mu,\qquad M_{\mu\nu}=M^\dagger_{\mu\nu}.
\eea
In terms of this realization, the functions $\phi^x_{\mu\nu}$ read
\be
\phi^x_{\mu\nu}=\eta_{\mu\nu}\varphi^x_1+\epsilon'\alpha^2x_\mu x_\nu\varphi^x_2,
\ee
and the Jacobi identities are satisfied if
\be\label{e14}
\psi^x=-2\varphi^x_1(\varphi^x_1)'+\varphi^x_1\varphi^x_2-2\epsilon'\alpha^2x^2(\varphi^x_1)'\varphi^x_2,
\ee
where now a prime denotes derivative with respect to $\epsilon'\alpha^2x^2$.
The function $\psi^x$ does not depend on $\chi^x$.

Since the Lorentz transformations are not deformed, the Casimir operator of this algebra is $\cC=x^2$.
The algebra (\ref{e10})-(\ref{e11}) is invariant under Hemitian conjugation if $\hp_\mu^\dagger=\hp_\mu$, $\hx_\mu^\dagger=\hx_\mu$,
$M_{\mu\nu}^\dagger=M_{\mu\nu}$.

Hermitian realizations of $\hp_\mu(\alpha)$ are obtained from (\ref{e12}) as
\be\label{e15}
\hp_\mu(\alpha)=\ha\big[p_\mu\varphi_1^x+\varphi_1^xp_\mu+\epsilon'\alpha^2(p{\cdot}x\,x_\mu\varphi^x_2+\varphi^x_2x_\mu x{\cdot}p)\big]+\epsilon'\alpha^2x_\mu\chi^x.
\ee


\section{Quantum deformed phase spaces depending on two parameters}

The deformed phase spaces of the previous sections can be unified by
introducing both fundamental Lorentz-invariant parameters $\alpha$ and $\beta$ so that \cite{10,11}
\be\label{e16}
[\hx_\mu,\hx_\nu]=i\epsilon\,\beta^2M_{\mu\nu}\;\psi^p(\epsilon\,\beta^2p^2),\qquad[\hp_\mu,\hp_\nu]=i\epsilon'\alpha^2M_{\mu\nu}\;\psi^x(\epsilon'\alpha^2x^2),\qquad[\hx_\mu,\hp_\nu]=ig_{\mu\nu},
\ee
where
\be\label{e17}
g_{\mu\nu}=h_0\eta_{\mu\nu}+\alpha^2X_{\mu\nu}+\beta^2P_{\mu\nu}+\alpha\beta H_{\mu\nu},
\ee
with $X_{\mu\nu}=h_1\hx_\mu\hx_\nu+\hx_\nu\hx_\mu h_1^\dagger$, $P_{\mu\nu}=h_2\hp_\mu\hp_\nu+\hp_\nu\hp_\mu h_2^\dagger$,
$H_{\mu\nu}=h_3(\hx_\mu\hp_\nu+\hp_\nu\hx_\mu)+h_4(\hx_\nu\hp_\mu+\hp_\mu\hx_\nu)+{\rm h.c.}$, and
the $h_i$ are Lorentz-invariant functions of $x_\mu$ and $p_\mu$, depending on $\hx^2$, $\hp^2$ and $D=\ha(\hx{\cdot}\hp+\hp{\cdot}\hx)$.

If $\hx_\mu$, $\hp_\mu$ and $M_{\mu\nu}$ are Hermitian operators, then $g^\dagger_{\mu\nu}=g_{\mu\nu}$, $h^\dagger=h$, $X^\dagger_{\mu\nu}=X_{\mu\nu}$,
$P^\dagger_{\mu\nu}=P_{\mu\nu}$ and $H^\dagger_{\mu\nu}=H_{\mu\nu}$.
The operators $g_{\mu\nu}$, $X_{\mu\nu}$, $P_{\mu\nu}$ and $H_{\mu\nu}$ transform as second rank tensors under Lorentz transformations, while $h_0$ is invariant,
$[M_{\mu\nu},h_0]=0$. We also require that all Jacobi identities hold. For simplicity, from now on we set $\epsilon=\epsilon'=1$.

Under suitable conditions on the $h_i$, the algebra (\ref{e16})-(\ref{e17}) satisfies the Born duality, defined as invariance for
$\beta\lra\alpha$, $\hx_\mu\rightarrow-\hp_\mu$, $\hp_\mu\rightarrow\hx_\mu$, $M_{\mu\nu}\lra M_{\mu\nu}$ and $g_{\mu\nu}\lra g_{\nu\mu}$ \cite{11}.
Moreover, it is easy to see that
\be
g_{\mu\nu}-g_{\nu\mu}=\alpha\beta(H_{\mu\nu}-H_{\nu\mu})=\alpha\beta FM_{\mu\nu},
\ee
with $[M_{\mu\nu},F]=0$.

The above quantum deformed phase spaces include as special cases Yang \cite{3} and TSR \cite{5,7} models and their generalizations \cite{4,11}.
The algebras with $X_{\mu\nu}=P_{\mu\nu}=0$ and $H_{\mu\nu}$ proportional to $M_{\mu\nu}$ reduce to Lie algebras, generated by $\hx_\mu$, $\hp_\mu$ $M_{\mu\nu}$ and $h_0$.
These algebras were introduced in \cite{4} and are defined by
\bea\label{e19}
&&[\hx_\mu,\hx_\nu]=i\beta^2M_{\mu\nu},\qquad[\hp_\mu,\hp_\nu]=i\alpha^2M_{\mu\nu},\qquad[\hx_\mu,\hp_\nu]=i(h_0\eta_{\mu\nu}-2\alpha\beta\rho M_{\mu\nu}),\cr
&&[h_0,\hx_\mu]=i(\beta^2\hp_\mu+2\alpha\beta\rho\,\hx_\mu),\qquad[h_0,\hp_\mu]=-i(\alpha^2\hx_\mu+2\alpha\beta\rho\,\hp_\mu),
\eea
where $\alpha$, $\beta$ and $\rho$ are real parameters. For $\rho=0$ one gets the Yang model, and for $\alpha=0$ or $\beta=0$ the Snyder or de Sitter algebra, respectively.

If instead $X_{\mu\nu}$ or $P_{\mu\nu}$ do not vanish, as in the case of TSR or SdS, the algebras  (\ref{e16})-(\ref{e17}) are not Lie algebras.
In the KL model $X_{\mu\nu}=P_{\mu\nu}=0$, but $H_{\mu\nu}$ is proportional to $M_{\mu\nu}$, and this gives rise to a Lie algebra.

\subsection{General Hermitian realizations}
The most general Hermitian realizations of the deformed phase spaces (\ref{e16})-(\ref{e17}) are given by
\be
\hx_\mu(\alpha,\beta)=e^{iG_1}\hx_\mu(\beta)\,e^{-iG_1},\qquad\hp_\mu(\alpha,\beta)=e^{iG_2}\hp_\mu(\alpha)\,e^{-iG_2},
\ee
where $\hx_\mu(\beta)$ and $\hp_\mu(\alpha)$ are given in (\ref{e9}) and (\ref{e15}) respectively, and
\be
G_1=G_1^{\,\dagger}=G_1(\alpha^2x^2,\alpha\beta D,\beta^2 p^2),\qquad G_2=G_2^{\,\dagger}=G_2(\alpha^2x^2,\alpha\beta D,\beta^2 p^2),
\ee
with $D=\ha(x{\cdot}p+p{\cdot} x)$.

In order for the Jacobi identities to hold, the functions $\varphi^p_1$, $\varphi^p_2$ and $\varphi^x_1$, $\varphi^x_2$ have to satisfy the relation (\ref{e7}) and (\ref{e14}) respectively.

In these realizations, the functions $X_{\mu\nu}$, $P_{\mu\nu}$, $H_{\mu\nu}$ and $h$ depend on $\varphi^p_1$, $\varphi^p_2$, $\varphi^x_1$, $\varphi^x_2$ and $G_1$, $G_2$.
The Born duality applied to the above realization generates new ones.
\subsection{Hermitian realization of the Yang model}
An Hermitian realization for the Yang model is given by
\bea
\hx_\mu(\beta)&=&\ha\left(x_\mu\sqrt{1-\beta^2p^2}+\sqrt{1-\beta^2p^2}\,x_\mu\right)+\beta^2p_\mu\chi^p(\beta^2p^2),\cr
\hp_\mu(\alpha)&=&\ha\left(p_\mu\sqrt{1-\alpha^2x^2}+\sqrt{1-\alpha^2x^2}\,p_\mu\right)+\alpha^2x_\mu\chi^x(\alpha^2x^2).
\eea

The Yang model is obtained for $G_1$, $G_2$ such that
\be
[\hx_\mu,\hp_\nu]=i\eta_{\mu\nu} h_0,\quad [h_0,\hx_\mu]=i\beta^2\hp_\mu, \quad [h_0,\hp_\mu]=-i\alpha^2\hx_\mu
\ee
hold. The explicit form at fourth order for $\chi^p=\chi^x=0$ is reported in \cite{11}.
In the limit $\alpha=\beta=0$, $\hx_\mu(\alpha,\beta)=x_\mu$ and $\hp_\mu(\alpha,\beta)=p_\mu$.

Similar constructions can be applied to TSR and SdS models, where
\bea&&\hx_\mu(\beta)=x_\mu+{\beta^2\over2}(\xp p_\mu+p_\mu\px)+\beta^2p_\mu\chi^p(\beta^2p^2),\cr
&&\hp_\mu(\alpha)=p_\mu+{\alpha^2\over2}(\px x_\mu+x_\mu\xp)+\alpha^2x_\mu\chi^x(\alpha^2x^2).
\eea

\section {Perturbative expansion of Hermitian realizations}
We shall now consider Hermitian realizations in a perturbative expansion in $\alpha$ and $\beta$ of the two-parameter model introduced in the previous section,
extending the results of \cite{11}.

\subsection{Second-order expansion}

At second order in $\alpha$ and $\beta$ we use the ansatz\footnote{In $\hx_\mu$ we could add also $\alpha\beta x_\mu$, $\alpha^2x_\mu x^2$, $\beta^2p_\mu$ and in
$\hp_\mu$ also $\alpha\beta p_\mu$, $\beta^2p_\mu p^2$, $\alpha^2x_\mu$.}
\be\label{e25}
\hx_\mu=x_\mu+{a_1\over2}\alpha\beta(x_\mu\xp+\px x_\mu)+{a_2\over2}\beta^2(x_\mu p^2+p^2 x_\mu)+{a_3\over2}\beta^2(p_\mu\px+\xp p_\mu)+{a_4\over2}\alpha\beta(p_\mu x^2+x^2p_\mu),
\ee
\be\label{e26}
\hp_\mu=p_\mu+{b_1\over2}\alpha\beta(p_\mu\px+\xp p_\mu)+{b_2\over2}\alpha^2(p_\mu x^2+x^2 p_\mu)+{b_3\over2}\alpha^2(x_\mu\xp+\px x_\mu)+{b_4\over2}\alpha\beta(x_\mu p^2+p^2x_\mu),
\ee
where $a_i$, $b_i$ are real constants.

Substituting (\ref{e25})-(\ref{e26}) in the relations $[\hx_\mu,\hx_\nu]=i\beta^2M_{\mu\nu}$ and $[\hp_\mu,\hp_\nu]=i\alpha^2M_{\mu\nu}$, we find $a_3-2a_2=1$, $b_3-2b_2=1$, while
calculating $[\hx_\mu,\hp_\nu]$  we get
\bea
g_{\mu\nu}&=&\ \eta_{\mu\nu}\left(1+2\tau\alpha\beta D+a_2\beta^2p^2+b_2\alpha^2x^2\right)+b_3\alpha^2x_\mu x_\nu+a_3\beta^2p_\mu p_\nu\cr
&&+\alpha\beta\Big[\tau(x_\mu p_\nu+p_\nu x_\mu)+\rho(x_\nu p_\mu+p_\mu x_\nu)\Big],
\eea
where $\rho=a_4+b_4$, $\tau=\ha(a_1+b_1)$, and $D=\ha(x{\cdot} p+p{\cdot} x)$.

Hence,
\bea
&&h_0=1+2\tau\alpha\beta D+a_2\beta^2p^2+b_2\alpha^2x^2,\qquad X_{\mu\nu}=b_3x_\mu x_\nu,\qquad P_{\mu\nu}=a_3p_\mu p_\nu,\cr
&&H_{\mu\nu}=\tau(x_\mu p_\nu+p_\nu x_\mu)+\rho(x_\nu p_\mu+p_\mu x_\nu).
\eea
In the following, for simplicity, we shall consider the symmetric solutions $a_i=b_i$.

For KL models, $a_2=b_2=-\ha$, $a_3=b_3=0$, $a_1=b_1=-\rho$, $a_4=b_4={\rho\over2}$, $\tau=-\rho$ and
\be
g_{\mu\nu}=\eta_{\mu\nu}\left(1-2\rho\alpha\beta D-\ha(\alpha^2x^2+\beta^2p^2)\right)-2\rho\alpha\beta M_{\mu\nu}=\eta_{\mu\nu} h_0-2\rho\alpha\beta M_{\mu\nu},
\ee
with $h_0=1-2\rho\alpha\beta D-\ha(\alpha^2x^2+\beta^2p^2)$,
and
\be
[h_0,\hx_\mu]=i(2\rho\alpha\beta x_\mu+\beta^2p_\mu),\qquad[h_0,\hp_\mu]=-i(2\rho\alpha\beta p_\mu+\alpha^2x_\mu).
\ee
For Yang models $\rho=0$ and $h_0=1-\ha(\alpha^2x^2+\beta^2p^2)$.
For TSR models $a_2=b_2=0$, $a_3=b_3=1$, $a_1=b_1=0$, $a_4=b_4=\ha$, $\tau=0$ and $\rho=1$.

\subsection{Fourth-order expansion}
At fourth order we use the ansatz\footnote{Also in this case we have omitted some terms that do not contribute significantly, like $\alpha^3\beta\,p_\mu x^4$,
$\alpha^4x_\mu x^4$ and $\beta^4p_\mu p^2$ in $\hx_\mu$ and analogous terms in $\hp_\mu$.}
\bea\label{e31}
\hx^{(4)}_\mu&=&\ {c_1\over2}\alpha^3\beta(x_\mu x^2\xp+\px x^2x_\mu)+{c_2\over2}\alpha^2\beta^2(x_\mu x^2p^2+p^2x^2x_\mu)+{c_3\over2}\alpha^2\beta^2(x_\mu\xp\px+\px\xp x_\mu)\cr
&&+{c_4\over2}\alpha\beta^3(x_\mu\xp p^2+p^2\px x_\mu)+{c_5\over2}\beta^4(x_\mu p^4+p^4x_\mu)+{c_6\over2}\beta^4(p_\mu p^2\px+\xp p^2p_\mu)\cr
&&+{c_7\over2}\alpha\beta^3(p_\mu p^2x^2+x^2p^2p_\mu)+{c_8\over2}\alpha\beta^3(p_\mu\px\xp+\px\xp p_\mu)\cr
&&+{c_9\over2}\alpha^2\beta^2(p_\mu\px x^2+x^2\xp p_\mu),
\eea
\bea\label{e32}
\hp^{(4)}_\mu&=&\ {d_1\over2}\alpha\beta^3(p_\mu p^2\px+\xp p^2p_\mu)+{d_2\over2}\alpha^2\beta^2(p_\mu p^2x^2+x^2p^2p_\mu)+{d_3\over2}\alpha^2\beta^2(p_\mu\px\xp+\xp\px p_\mu)&\cr
&&+{d_4\over2}\alpha^3\beta(p_\mu\px x^2+x^2\xp p_\mu)+{d_5\over2}\alpha^4(p_\mu x^4+x^4p_\mu)+{d_6\over2}\alpha^4(x_\mu x^2\xp+\px x^2x_\mu)&\cr
&&+{d_7\over2}\alpha^3\beta(x_\mu x^2p^2+p^2x^2x_\mu)+{d_8\over2}\alpha^3\beta(x_\mu\xp\px+\xp\px x_\mu)\cr
&&+{d_9\over2}\alpha^2\beta^2(x_\mu\xp p^2+p^2\px x_\mu),
\eea
where $c_i$, $d_i$ are real constants.

Inserting (\ref{e31}) and (\ref{e32}) into the relations $[\hx_\mu,\hx_\nu]=i\beta^2M_{\mu\nu}$ and $[\hp_\mu,\hp_\nu]=i\alpha^2M_{\mu\nu}$, we find
$c_6=4a_2^2+4c_5+a_2$, $d_6=4b_2^2+4d_5+b_2$.

For KL models, assuming $d_i=c_i$, one has $c_1=c_4=c_6=c_8=0$, $c_2={\rho^2\over8}$, $c_3=\qu-{\rho^2\over2}$, $c_5=-{1\over8}$, $c_7={\rho\over4}$,
$c_9={\rho^2\over4}$, and calculating $[\hx_\mu,\hp_\nu]$ one obtains
\bea
h_0&=&\ 1-2\rho\alpha\beta D-\ha(\alpha^2x^2+\beta^2p^2)-{1\over8}(\alpha^4x^4+\beta^4p^4)+{\rho\over4}\alpha^3\beta(x^2x{\cdot} p+p{\cdot} x\,x^2)\cr
&&+{\rho\over4}\alpha\beta^3(p^2p{\cdot} x+x{\cdot} p\,p^2)+{1\over8}\alpha^2\beta^2(x^2p^2+p^2x^2)+\ha\alpha^2\beta^2D^2,
\eea
and
\bea
[h_0,\hx_\mu]&=&\ i\Big[2\rho\alpha\beta x_\mu+\beta^2p_\mu+\left({\rho^2\over2}-\qu\right)\alpha^2\beta^2(x^2p_\mu+p_\mu x^2)-{\rho\over2}\alpha\beta^3(\xp p_\mu+p_\mu \px)\cr
&&-\rho^2\alpha^2\beta^2(\px x_\mu+x_\mu\xp)\Big]=i(\beta^2\hp_\mu+2\alpha\beta\rho\,\hx_\mu),
\eea
\bea
[h_0,\hp_\mu]=&&-i\Big[2\rho\alpha\beta p_\mu+\alpha^2x_\mu+\left({\rho^2\over2}-\qu\right)\alpha^2\beta^2(p^2x_\mu+x_\mu p^2)-{\rho\over2}\alpha^3\beta(\px x_\mu+x_\mu \xp)\cr
&&-\rho^2\alpha^2\beta^2(\px p_\mu+p_\mu\px)\Big]=-i(\alpha^2\hx_\mu+2\alpha\beta\rho\,\hp_\mu)
\eea
in accordance with (\ref{e19}).
The Yang model is obtained for $\rho=0$.

For SdS, the results are reported in  \cite{11}.
In particular, for $c_i=d_i$, the coefficients depend on the free parameter $c_1$, with $c_5=c_6=c_7=0$, $c_2={1\over8}$, $c_3=\ha$, $c_4=\ha-c_1$,
$c_8=1-c_1$, $c_9={3\over4}$.

\section{Exact results on generalized TSR}
In this section, we present some exact realizations of generalized TSR obtained by exploiting a method proposed in \cite{7,8}.

Let us start with the Snyder algebra
\be
[\hx_\mu,\hx_\nu]=i\beta^2M_{\mu\nu},\qquad \beta\ne0.
\ee
A class of realizations of $\hx_\mu$ is given by
\be
\hx_\mu=X_\mu\varphi_1(\beta^2P^2)+\beta^2X{\cdot} P\,P_\mu\varphi_2(\beta^2P^2),\qquad M_{\mu\nu}=X_\mu P_\nu-X_\nu P_\mu=M_{\mu\nu}^\dagger,
\ee
with $\hx^\dagger\ne\hx$, where $X_\mu$ and $P_\mu$ satisfy
\be
[X_\mu,X_\nu]=[P_\mu,P_\nu]=0,\qquad [X_\mu, P_\nu]=i\eta_{\mu\nu},
\ee
and
\be
\varphi_2={1+2\varphi_1\varphi'_1\over\varphi_1-2\beta^2P^2\varphi_1'},\qquad{\rm with}\ \varphi'_1={d\varphi_1\over d(\beta^2P^2)},\quad \varphi(0)=1.
\ee

Let us define
\be
\hp_\mu=P_\mu-\epsilon{\alpha\over\beta}\hx_\mu,\qquad \epsilon^2=1,\ \alpha\ne0,
\ee
and using $[\hx_\mu,P_\nu]=[\hx_\nu,P_\mu]$ we obtain
\be
[\hp_\mu,\hp_\nu]=i\alpha^2M_{\mu\nu}, \qquad[\hx_\mu,\hp_\nu]=ig_{\mu\nu}=i\eta_{\mu\nu}\varphi_1(\beta^2P^2)+\beta^2P_\mu P_\nu\varphi_2(\beta^2P^2)-\epsilon\,\alpha\beta M_{\mu\nu}.
\ee
Hence, $g_{\mu\nu}-g_{\nu\mu}=-2\epsilon\,\alpha\beta M_{\mu\nu}$.

Note that
\be
P_\mu=\hp_\mu+\epsilon{\alpha\over\beta}\hx_\mu,
\ee
and
\be
M_{\mu\nu}=X_\mu P_\nu-X_\nu P_\mu=(\hx_\mu P_\nu-\hx_\nu P_\mu){1\over\varphi_1(\beta^2P^2)}=\left(\hx_\mu\hp_\nu-\hx_\nu\hp_\mu-2i\epsilon\,\alpha\beta M_{\mu\nu}\right){1\over\varphi_1(\beta^2P^2)}.
\ee
Hence,
\be
M_{\mu\nu}=\ha(\hx_\mu\hp_\nu-\hx_\nu\hp_\mu+\hp_\nu\hx_\mu-\hp_\mu\hx_\nu){1\over\varphi_1(\beta^2P^2)}.
\ee

The algebra generated by $\hx_\mu$, $\hp_\mu$ and $M_{\mu\nu}$ and all its realizations are invariant under the Born duality $\alpha\lra\beta$,
$\hx_\mu\rightarrow-\hp_\mu$, $\hp_\mu\rightarrow \hx_\mu$, $M_{\mu\nu}\lra M_{\mu\nu}$ and $g_{\mu\nu}\lra g_{\nu\mu}$. The relation between $\hx_\mu$ and $\hp_\mu$ can be written as
\be
\hp_\mu=-\epsilon{\alpha\over\beta}\,S\hx_\mu S^{-1},
\ee
where
\be
S=\exp\left(-{iZ\over2\epsilon\,\alpha\beta}\right),\qquad{dZ\over d(\beta^2P^2)}={1\over\varphi_1(\beta^2P^2)+\beta^2P^2\varphi_2(\beta^2P^2)}.
\ee
Note that $Z^\dagger=Z$, $S^{-1}=S^\dagger$.

Clearly, Hermitian realizations of $\hx_\mu$ and $\hp_\mu$ are given by
\be
\hx_\mu^H=\ha(\hx_\mu+\hx_\mu^\dagger),\qquad\hp_\mu^H=\ha(\hp_\mu+\hp_\mu^\dagger).
\ee

\subsection{Special cases}
For $\varphi_1=\varphi_2=1$, we obtain SdS \cite{7,11}
\bea
&&[\hx_\mu,\hp_\nu]=i(\eta_{\mu\nu}+\beta^2P_\mu P_\nu-\epsilon\alpha\beta M_{\mu\nu})=i\big(\eta_{\mu\nu}
+(\alpha\hx_\mu+\beta\hp_\mu)(\alpha\hx_\nu+\beta\hp_\nu) -\epsilon\alpha\beta M_{\mu\nu}\big),\cr
&&M_{\mu\nu}=\ha\left(\hx_\mu\hp_\nu-\hx_\nu\hp_\mu-\hp_\mu\hx_\nu+\hp_\nu\hx_\mu\right).
\eea
In this case,
\be
Z=\ln(1+\beta^2P^2),\qquad S=\exp\left(-{i\over2\epsilon\alpha\beta}\ln(1+\beta^2P^2)\right).
\ee
\medskip

For $\varphi_1=\sqrt{1-\beta^2P^2}$, $\varphi_2=0$, we obtain
\bea
&&[\hx_\mu,\hp_\nu]=i\left(\eta_{\mu\nu}\sqrt{1-\beta^2P^2}-\epsilon\alpha\beta M_{\mu\nu}\right),\cr
&&M_{\mu\nu}=\ha\left(\hx_\mu\hp_\nu-\hx_\nu\hp_\mu-\hp_\mu\hx_\nu+\hp_\nu\hx_\mu\right){1\over\sqrt{1-\beta^2P^2}}.
\eea
In this case,
\be
Z=-2\sqrt{1-\beta^2P^2},\qquad S=\exp\left({i\over\epsilon\alpha\beta}\sqrt{1-\beta^2P^2}\right).
\ee
This example corresponds to a special case of KL model \cite{4} with
$h_0=\sqrt{1-\beta^2P^2}$, $\epsilon=2\rho=\pm1$, and
\be
[h_0,\hx_\mu]=i(\beta^2\hp_\mu+\epsilon\,\alpha\beta\hx_\mu),\qquad[h_0,\hp_\mu]=-i(\alpha^2\hx_\mu+\epsilon\,\alpha\beta\hp_\mu).
\ee
\section{Conclusions}
Models of noncommutative geometry in curved spacetime have recently attracted much interest because of
possible applications to astrophysical observations \cite{26} and to the measurement of time delay in the propagation
of photons by cosmic sources.

In this paper we have examined a class of these models, characterized by a high degree of symmetry, which
generalize an early proposal by Yang \cite{3}, and include  TSR \cite{5,7} among others.
The main feature of these models is that the defining algebra contains both the Snyder \cite{2} and the de Sitter \cite{23} algebra,
and in particular, the Lorentz invariance is preserved. They may therefore be relevant in a low-energy limit
of quantum gravity, for which theoretical arguments suggest that both a noncommutative parameter and a cosmological constant
should be relevant \cite{5}.

We have presented quantum realizations of these algebras in canonical phase space, starting from the simpler cases
of Snyder or its de Sitter dual.
However, their structure, that involves the full phase space, renders problematic the definition of a star product or of a Hopf algebroid
structure, like those introduced in \cite{27}.
More general mathematical constructions should be introduced if one wishes to include analogous notions in this formalism.

A possible area of application of our results is quantum field theory. A field theory based on the SdS algebra was discussed in \cite{20},
based on rough approximations. In that paper it was noted its similitude with the Grosse-Wulkenhaar model \cite{28}, a renormalizable and exactly
solvable theory, which, in analogy with SdS field theory, can be interpreted as a field theory in curved noncommutative space \cite{29}.
An investigation of the field theory on spacetimes of Yang type that exploit the general results obtained in this paper would therefore be a
promising development of the present research.

\section{Ackowledgements}
S. Mignemi acknowledges support from GNFM and COST action CA18108.

\end{document}